\documentclass[aps,prl,preprint,groupedaddress]{revtex4}
\usepackage{graphicx} 
\usepackage{dcolumn}
\usepackage{bm}
\begin{document}

\title{High-temperature magnetic properties of noninteracting single-domain Fe$_{3}$O$_{4}$
nanoparticles} 
\author{Jun Wang$^{1,a}$, Pieder Beeli$^{2}$, L. H. Meng$^{1}$, and Guo-meng Zhao$^{1,2,b}$} 
\affiliation{
$^{1}$Department of Physics, Faculty of Science, Ningbo
University, Ningbo, P. R. China~\\
$^{2}$Department of Physics and Astronomy, 
California State University, Los Angeles, CA 90032, USA}

\begin{abstract}

 Magnetic measurements have been performed on 40-nm sphere-like Fe$_{3}$O$_{4}$ nanoparticles using a Quantum
 Design vibrating sample magnetometer. Coating
 Fe$_{3}$O$_{4}$ nanoparticles with SiO$_{2}$ effectively
eliminates magnetic interparticle interactions so that
the coercive field $H_{C}$ in the high-temperature range between 300 K and 
the Curie temperature (855 K) can be well fitted by an expression for noninteracting randomly oriented
single-domain particles. From the fitting parameters, the effective anisotropy
constant $K$ is found to be (1.68$\pm$0.17)$\times$10$^{5}$ erg/cm$^{3}$, which is 
slightly larger than the
bulk magnetocrystalline anisotropy constant of 1.35$\times$10$^{5}$ erg/cm$^{3}$.
Moreover, the inferred mean particle diameter from the fitting
parameters is in quantitative agreement with that determined from transmission electron microscope. 
Such a quantitative agreement between data and theory suggests that the assemble of 
our SiO$_{2}$-coated sphere-like Fe$_{3}$O$_{4}$
nanopartles represents a good system of noninteracting
randomly-oriented single-domain particles.

\end{abstract}
\maketitle 

Ensembles of magnetic nanoparticles in various forms
have been at the focus of scientific interest \cite{book} since the days of N\'eel
\cite{Neel}
and Brown \cite{Brown}, who developed a theory for noninteracting single-domain ferromagnetic particles.  A 
complete understanding of the 
magnetic properties
of ferromagnetic nanoparticles is not simple, in particular because of
the complexity of real nanoparticle assemblies, involving magnetic 
interparticle interactions and magnetic anisotropy. 
An important contribution to the understanding of the
magnetic behavior of nanoparticles was given by Bean and
Livingston (BL) \cite{BL} who assumed an assembly of noninteracting single-domain
particles with uniaxial anisotropy. This study was
based on the N\'eel relaxation time $\tau =\tau_{0}\exp (KV/k_{B}T)$, 
where $\tau_{0}$ is the characteristic time constant, $k_{B}$ is the Boltzmann constant, $K$ is
the uniaxial anisotropy constant, and $V$ is the particle volume.
$KV$ represents the energy barrier between two easy directions.
According to Bean and Livingston, at a given observation
time $\tau_{obs}$, there is a critical temperature, called the
blocking temperature $T_{B}$, given by \cite{BL}

\begin{equation}
    T_{B} = \frac{KV}{k_{B}\ln (\tau_{obs}/\tau_{0})},
    \end{equation}
above which the magnetization reversal of an assembly of
identical single-domain particles goes from
blocked (having hysteresis) to superparamagnetic-type behavior. Within this 
framework the coercive field $H_{C}$ is expected to decrease
with the square root of temperature:

\begin{equation}
    H_{C} = \alpha\frac{2K}{M_{s}}[1-(T/T_{B})^{1/2}],
    \end{equation}
where $M_{s}$ is the zero-temperature saturation magnetization and $\alpha$ = 1 if the
particle easy-axes are aligned \cite{BL} or $\alpha$ = 0.48 if randomly
oriented \cite{Kech}.

\begin{figure}[htb]
     \vspace{-0.2cm}
    \includegraphics[height=6cm]{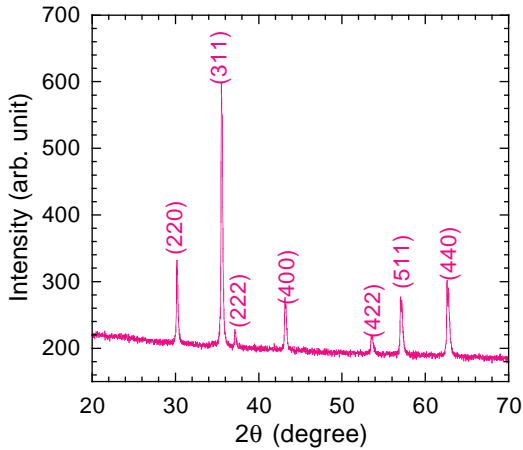}
     \vspace{-0.3cm}
 \caption[~]{ X-ray diffraction (XRD) spectrum of  hydrothermally 
 synthesized Fe$_{3}$O$_{4}$ nanoparticles. } 
\end{figure}

The above equations have not been well tested by experiments due to
the experimental difficulties in producing assemblies of
noninteracting sphere-like nanoparticles.
When magnetic nanoparticles are closely packed and/or aggregate, the interparticle interactions 
are expected to modify the magnetic behavior of the assembly. These interactions
can have a dipolar, Ruderman-Kittel-Kasuya-Yosida
(RKKY), or a superexchange character, depending on the character of an assemble. 
For magnetic nanoparticles embedded in an insulating matrix such as amorphous alumina 
\cite{Dorm88} and amorphous SiO$_{2}$ \cite{Aliev,Mitra,Yang}),
the dipolar interactions are the dominant ones~\cite{Kech}. Major theoretical
and experimental efforts have been focused on
the understanding of the role of the dipolar interactions \cite{Dorm}.
 In addition to granular metal solids \cite{Dorm88,Aliev,Mitra,Yang,Chien}, frozen ferrofluids have been used to 
 investigate the role of dipolar interactions \cite{Chan,Hilo,Luo,Moi}. In these systems, the 
 magnetic particles
 are held fixed in a frozen insulating
liquid. The degree of dilution in the liquid solvent controls
the average particle distance and therefore the strength of the interactions.
However, these studies have been limited to ultra-fine particles with 
a low $T_{B}$ and to a temperature region well below the Curie temperature ($T_{C}$) of the magnetic 
nanoparticles. Since the particles are so fine, the contribution of 
the surface anisotropy becomes significant and even dominant if they
are not perfectly spherical particles \cite{Bod}.

\begin{figure}[htb]
     \vspace{-0.2cm}
    \includegraphics[height=14cm]{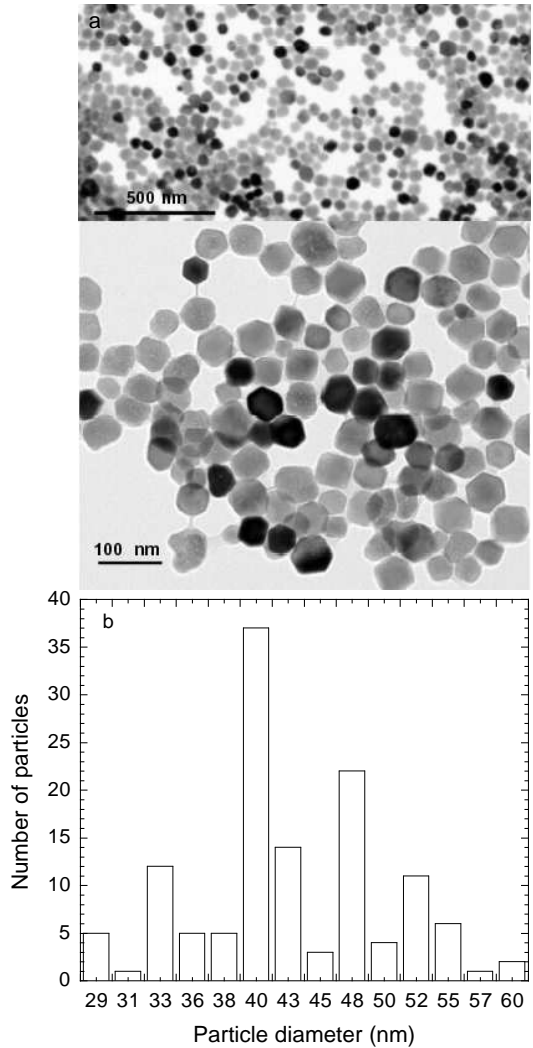}
     \vspace{-0.3cm}
 \caption[~]{a) Transmission electron microscopy images of 
 as grown Fe$_{3}$O$_{4}$ nanoparticles. b) Histograph of the particle size distribution.
 The mean
diameter of the particles is found to be about 43 nm.} 
\end{figure}

Here we report magnetic measurements on
40-nm sphere-like Fe$_{3}$O$_{4}$ nanoparticles using a Quantum
 Design vibrating sample magnetometer. Coating
 Fe$_{3}$O$_{4}$ nanoparticles with SiO$_{2}$ effectively
eliminates magnetic interparticle interactions so that
the coercive field $H_{C}$ in the high-temperature range between 300 K and 
$T_{C}$ follows Eq.~(2) for noninteracting randomly oriented
particles. Fitting the data with Eq.~(2) yields 
$K$ = (1.68$\pm$0.17)$\times$10$^{5}$ erg/cm$^{3}$, which is slightly
larger than the bulk magnetocrystalline anisotropy constant of 1.35$\times$10$^{5}$ erg/cm$^{3}$
(Ref.~\cite{Cull}). Moreover, the inferred mean particle diameter from the fitting parameters
is in good agreement with that determined from transmission electron microscope.
Such a good agreement between data and theory suggests that the assemble of our 
SiO$_{2}$-coated sphere-like Fe$_{3}$O$_{4}$
nanopartles represents a good system of noninteracting
randomly-oriented single-domain particles.

Samples were synthesized by an improved hydrothermal route \cite{Wang2} using the following reagents and solvents: iron(III) chloride 
hexahydrate, diethylene glycol,
sodium hydroxide, and iron(II) chloride tetrahydrate.  The as-grown Fe$_{3}$O$_{4}$ nanoparticles
were then coated  with SiO$_{2}$ following the method of Ref.~\cite{Wang3}.
Magnetization was measured using a Quantum Design vibrating sample magnetometer (VSM).
The moment measurement was carried out after the sample chamber reached a high vacuum of 
better than 9$\times$10$^{-6}$ torr. The absolute measurement
uncertainty in moment is less than 1$\times$10$^{-6}$ emu.

Figure~1 shows x-ray diffraction (XRD) spectrum of  hydrothermally 
 synthesized Fe$_{3}$O$_{4}$ nanoparticles. The positions and relative
intensities of all diffraction peaks match well with those of
JCPDS card (19-0629) of magnetite with a lattice constant of 8.367~\AA.

 Figure~2a shows Transmission electron microscopy (TEM) images of 
 as synthesized Fe$_{3}$O$_{4}$ nanoparticles.
 The TEM images were taken with a Hitachi model H-800 using an accelerating voltage of 
 80 kV. The pictures demonstrate high quality and
monodispersity of the as synthesized nanoparticles.  Fig.~2b displays
a histograph of the particle size distribution. From the histograph,
we determine the mean
diameter of the particles to be about 43 nm.

\begin{figure}[htb]
     \vspace{-0.2cm}
    \includegraphics[height=12cm]{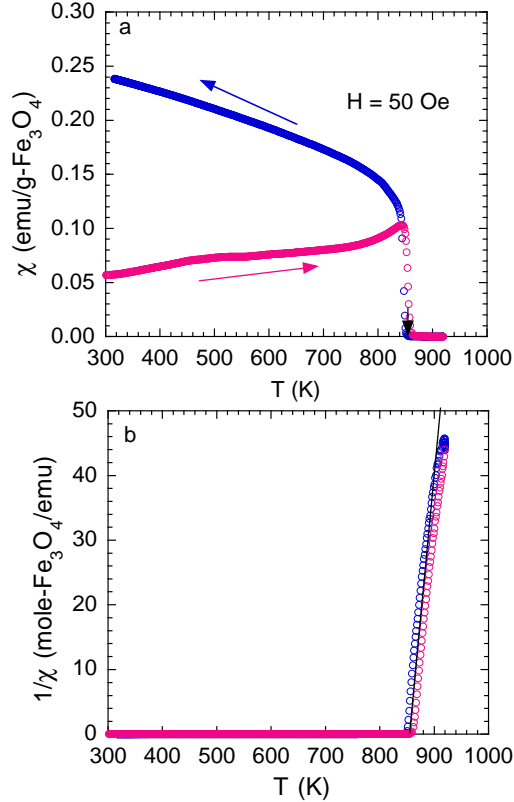}
     \vspace{-0.6cm}
 \caption[~]{a) Temperature dependencies of the ZFC and FC
 susceptibilities  for
 Fe$_{3}$O$_{4}$ nanoparticles coated with SiO$_{2}$, which were measured 
 in a field of 50 Oe. The downward arrow indicates a Curie temperature $T_{C}$ of 855 K.
 b) The reciprocal of the susceptibility $1/\chi$
versus temperature. The Curie-Weiss fit (solid line) yields $T_{C}$ = 
855 K. } 
\end{figure}

Figure~3a shows temperature dependencies of the ZFC and FC
 susceptibilities for Fe$_{3}$O$_{4}$ nanoparticles coated with SiO$_{2}$, which 
 were measured in a field of 50 Oe.  The susceptibility was
 calculated using the fact that the SiO$_{2}$-coated sample
contains 75$\%$ Fe$_{3}$O$_{4}$ and 25$\%$ SiO$_{2}$ (in weight), which
were determined from the measured
room-temperature saturation 
magnetizations of both as-grown and SiO$_{2}$-coated Fe$_{3}$O$_{4}$ samples. One can clearly 
see that there are significant
 differences between the ZFC and FC susceptibilities in the whole
 temperature range between 300 K and  $T_{C}$. This indicates
that the blocking temperature of the nanoparticle assembly is higher
than $T_{C}$. A Curie
temperature of 855 K is inferred from the data (indicated by the
downward arrow), which takes into
account a small thermal lag. In Fig.~3b, we plot  the reciprocal of the susceptibility $1/\chi$
versus temperature. It is apparent that the susceptibility data 
above the Curie temperature can be well fitted by 
the Curie-Weiss law (solid line): $\chi = C/(T - T_{C})$ with $T_{C}$ 
= 855 K and the Curie-Weiss constant $C$ = 1.12 emu/K mole-Fe$_{3}$O$_{4}$. 
It is remarkable that the $T_{C}$'s determined from the data above and
below $T_{C}$ are almost identical.
The value of the Curie-Weiss constant corresponds to an effecctive moment $p_{eff}$
= 3.0 $\mu_{B}$ per Fe$_{3}$O$_{4}$ (where $\mu_{B}$ is the Bohr magneton).

\begin{figure}[htb]
     \vspace{-0.2cm}
    \includegraphics[height=13cm]{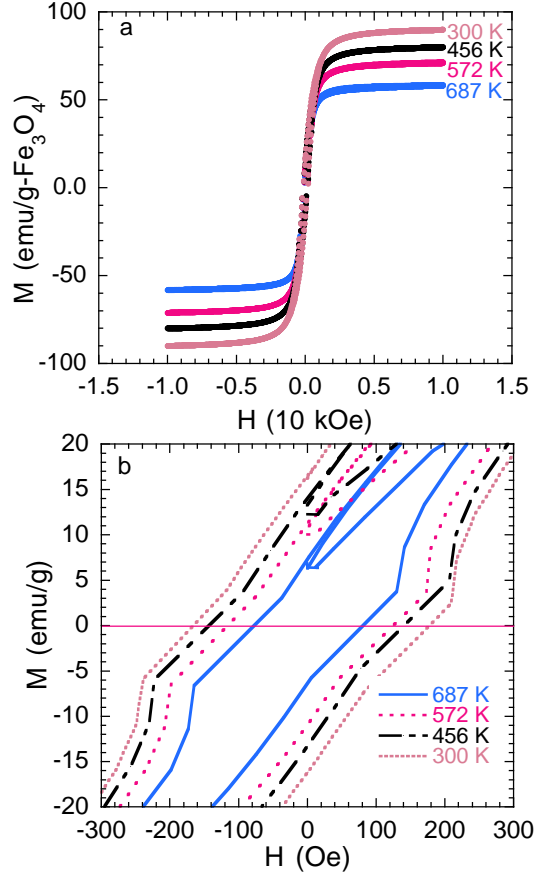}
     \vspace{-0.6cm}
 \caption[~]{a) Magnetic hysteresis loops at four different
temperatures. b) An expanded view of magnetic hysteresis loops at four different
temperatures.} 
\end{figure} 

In Fig.~4a we plot magnetic hysteresis loops at  four different
temperatures.  The magnetization at $H$ = 10 kOe progressively
decreases as the temperature increases and is
almost saturated at 10 kOe.    In order to see the low field data more clearly, we show an expanded
view of the loops in Fig.~4b. It is clear that the cocercive field $H_{C}$ also decreases
progressively as the temperature increases.

Figure~5a shows the saturation magnetization $M_{s}$ as a function of
temperature. The $M_{s}$ value at $T_{C}$ = 855 K is set to zero. The 
solid line is a fitted curve by $M_{s} = A (T-T_{C})^{\beta}$ with $\beta$
= 0.354. The $M_{s}$ at room temperature is 86.6
emu/g-Fe$_{3}$O$_{4}$, which is slightly lower than the bulk value of 
92 emu/g-Fe$_{3}$O$_{4}$. Then the zero-temperature $M_{s}$ should be about 90 emu/g-Fe$_{3}$O$_{4}$.

\begin{figure}[htb]
     \vspace{-0.2cm}
    \includegraphics[height=12cm]{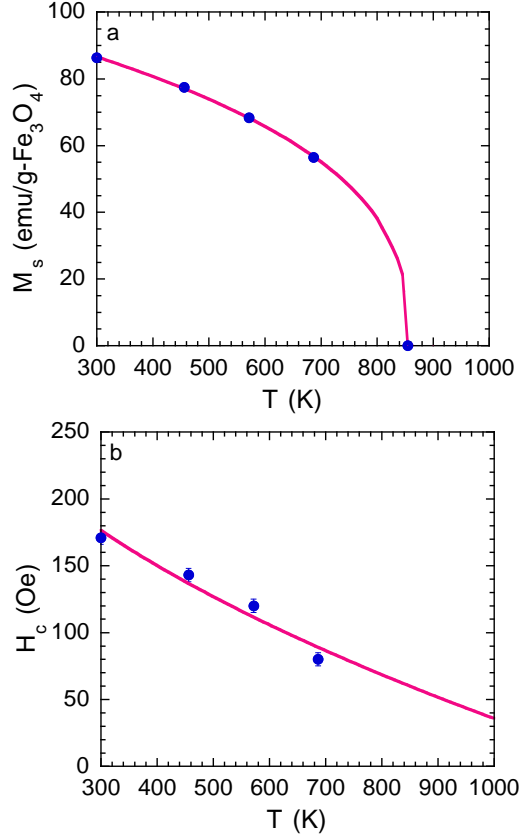}
     \vspace{-0.3cm}
 \caption[~]{a) Temperature dependence of the saturation magnetization $M_{s}$.
 The solid line is a fitted curve by $M_{s} = A (T-T_{C})^{\beta}$ with $\beta$
= 0.354 and $T_{C}$ = 855 K.
 b) Temperature dependence of the cocercive field $H_{C}$. The solid
 line is the fitted curve by Eq.~(2) with the fitting
parameters: $T_{B}$ = 1243$\pm$153 K and $K/M_{s}$ = 361$\pm$37 Oe. } 
\end{figure} 

In Fig.~5b, we plot $H_{C}$ versus $T$ for the sample. Since the $H_{C}$
values are found to be slightly different from the positive and negative field data due
to a remanent magnetic field of about 10 Oe in the superconducting
magnet of the equipment, we take $H_{C}$ to be the average of the two $H_{C}$
values obtained from the positive and negative
field data, respectively. The solid line is a fitted curve by Eq.~(2) with the fitting
parameters $T_{B}$ = 1243$\pm$153 K and $K/M_{s}$ = 361$\pm$37 Oe.
Using $M_{s}$ = 90 emu/g and $K/M_{s}$ = 361$\pm$37 Oe, we find
$K$ = (1.68$\pm$0.17)$\times$10$^{5}$ erg/cm$^{3}$, which is slightly larger than the
bulk value of 1.35$\times$10$^{5}$ erg/cm$^{3}$ (Ref.~\cite{Cull}). If
we use $M_{s}$ = 77.8 emu/g (Ref.~\cite{Goya}), we obtain 
$K$ = (1.45$\pm$0.14)$\times$10$^{5}$ erg/cm$^{3}$, which is very
close to the bulk value.

In order to further check whether our data are in quantitative agreement with 
the theoretical predictions [Eqs.~(1) and (2)], we use
Eq.~(1), the bulk $K$, and the inferred $T_{B}$ to
estimate the average particle diameter. Since the average measuring time for each data point is 0.5
s, we can set $\tau_{obs}$ = 0.5 s. The value of $\tau_{0}$ for Fe$_{3}$O$_{4}$
nanoparticles was found \cite{Goya}
to be 9$\times$10$^{-13}$ s. Substituting $\tau_{obs}$ = 0.5 s, $\tau_{0}$
= 9$\times$10$^{-13}$ s, and $T_{B}$ = 1243$\pm$153 K
into Eq.~(1), we find $d$ = 40.4$\pm$2.1 nm, which is close to that
(43$\pm$2 nm) deduced from TEM. Using the diameter of 40 nm for as-grown Fe$_{3}$O$_{4}$
nanoparticles and 
the fact that the SiO$_{2}$-coated sample contains 75$\%$ Fe$_{3}$O$_{4}$ and 25$\%$
SiO$_{2}$ (in weight), we estimate the average thickness of the coated SiO$_{2}$ layers to be about 
6 nm.

The quantitative agreement between our data and the BL theory suggest that the
assembly of our 40-nm Fe$_{3}$O$_{4}$ nanoparticles coated with 6-nm
SiO$_{2}$ layers 
respresents a nearly ideal system of noninteracting single-domain
particles. The diameter of 40 nm is well
below the maximum diameter of single-domain particles of about 128 nm 
(Refs~\cite{Sor,Kit}). The  fact that the inferred 
$K$ value is close to the bulk one suggests that a contribution of 
the surface anisotropy is small, in agreement with the
sphere-like shape of particles (see the TEM pictures in Fig.~2a).

In summary, we have made high-temperature magnetic measurements on hydrothermally 
 synthesized Fe$_{3}$O$_{4}$ nanoparticles using a Quantum
 Design vibrating sample magnetometer. Coating
 40-nm Fe$_{3}$O$_{4}$ nanoparticles with about 6-nm SiO$_{2}$ effectively
reduces magnetic interparticle interactions so that
the coercive field $H_{C}$ follows the BL expression for noninteracting
single-domain magnetic particles. The quantitative agreement 
between our data and the BL theory \cite{BL} suggests that the assemble of our SiO$_{2}$-coated 
sphere-like Fe$_{3}$O$_{4}$
nanopartles represents a nearly ideal system of noninteracting
randomly-oriented single-domain particles.

{\bf Acknowledgment:}
This work was supported by the National Natural Science Foundation of China (10874095), 
the Science Foundation of China, Zhejiang (Y407267, 2009C31149), the Natural Science Foundation of Ningbo 
(2008B10051, 2009B21003), K. C. Wong Magna Foundation, and Y. G. Bao's Foundation. 

~\\
~\\
$^{a}$ wangjun2@nbu.edu.cn~\\
$^{b}$ gzhao2@calstatela.edu

\bibliographystyle{prsty}

\end{document}